\documentclass[12pt]{article}
\usepackage{amsmath,amssymb,amsthm,amsfonts,latexsym}

\usepackage{graphicx}
\usepackage{rotating}
\usepackage{longtable}
\usepackage{booktabs}
\usepackage{array}
\usepackage{dcolumn}
\usepackage{multirow}
\usepackage[flushleft]{threeparttable} 
\usepackage{float}
\usepackage{wrapfig}
\usepackage{booktabs,tabularx}
\usepackage{pifont}
\usepackage{placeins}
\usepackage{natbib}
\usepackage{hyperref}
\usepackage{listings}
\usepackage{xcolor}
\usepackage{fancyvrb}
\usepackage{authblk}

\newcommand{\pkg}[1]{\textbf{\texttt{#1}}}

\usepackage{enumitem}
\usepackage{units}
\usepackage{esint}

\usepackage{xcolor}

\usepackage{tikz,pgfplots}
\pgfplotsset{compat=1.17}
\usetikzlibrary{
  shapes.geometric,
  shapes.misc,
  positioning,
  chains,matrix,scopes,
  arrows,decorations.markings,arrows.meta,
  decorations.pathmorphing,
  shadows
}

\def\simind{\stackrel{ind}{\sim}}   

\newcommand{\BF}{\boldsymbol}

\begin{document}

\setlength{\baselineskip}{23pt}

\title{
    mmcmcBayes: An R Package Implementing a Multistage MCMC Framework for Detecting the Differentially Methylated Regions
	\vskip1em
}
\author[1]{Zhexuan Yang}
\author[2]{Duchwan Ryu}
\author[2]{Feng Luan}

\affil[1]{Department of Statistics, The Pennsylvania State University}

\affil[2]{Department of Statistics and Actuarial Science, Northern Illinois University}

\date{}
\maketitle

\begin{center}
    \textbf{Abstract}
\end{center}
    
Identifying differentially methylated regions is an important task in epigenome-wide association studies, where differential signals often arise across groups of neighboring CpG sites. Many existing methods detect differentially methylated regions by aggregating CpG-level test results, which may limit their ability to capture complex regional methylation patterns. In this paper, we introduce the \texttt{R} package \textbf{mmcmcBayes}, which implements a multistage Markov chain Monte Carlo procedure for region-level detection of differentially methylated regions. The method models sample-wise regional methylation summaries using the alpha-skew generalized normal distribution and evaluates evidence for differential methylation between groups through Bayes factors. We use a multistage region-splitting strategy to refine candidate regions based on statistical evidence. We describe the underlying methodology and software implementation, and illustrate its performance through simulation studies and applications to Illumina 450K methylation data. The \textbf{mmcmcBayes} package provides a practical region-level alternative to existing CpG-based differentially methylated regions detection methods and includes supporting functions for summarizing, comparing, and visualizing detected regions. 

\vspace{0.5cm}


\section{Introduction}
\label{sec:intro} 
The genome stores epigenetic information through heritable modifications to DNA's chemical structure, such as the methylation of specific bases. One form of chemical modification of DNA is known as DNA methylation. DNA methylation is a fundamental epigenetic mechanism involved in gene regulation, cellular differentiation, and disease development. In human cells, approximately one percent of DNA bases are methylated, with methylation occurring predominantly at cytosine-guanine phosphate (CpG) sites in mature somatic tissues. 

While early analyses focused on individual CpG sites, it is now well recognized that biologically meaningful signals often occur at the regional level, leading to increased interest in the identification of differentially methylated regions (DMRs).
As \cite{Rakyan11} pointed out, the DMR is defined as a region of the genome at which multiple adjacent CpG sites show differential methylation, and DMR is recognized as a functional region with the potential to play a role in the control of the regulation of gene transcription levels.

DMRs have been reported in a wide range of disease related studies, including 
Down syndrome \citep{Jin13}, Alzheimer's disease \citep{Zhang2020}, rheumatoid arthritis \citep{Shao2019}, diabetes \citep{Johnson2020, Dayeh2014}, systemic lupus erythematosus \citep{Zhang2023}, autoimmune thyroid disease \citep{Lafontaine2024, Absher2013, Liu2013}, neurodevelopmental disorders \citep{LaddAcosta2014},  and cancer \citep{Irizarry2009, Hansen2011}
These findings highlight the importance of region-level inference for understanding epigenetic variation associated with disease processes. From a statistical perspective, however, the detection of DMRs remains challenging due to the high dimensionality of methylation data, strong spatial correlation among CpG sites, and heterogeneous distributional patterns across samples. 

Most existing methods for DMR detection operate by first performing inference at the CpG level and then aggregating site-wise results to define regions. For example, 
the package \textbf{DMRcate} \citep{Peters2015} applies kernel smoothing to CpG-level test statistics, while \textbf{bumphunter} \citep{AndrewJaffe12} identifies contiguous genomic segments where smoothed regression coefficients exceed predefined thresholds. Similarly, 
\textbf{DMRcaller} \citep{Catoni2018}, \textbf{DSS} \citep{Feng2014}, and \textbf{bsseq/BSmooth} \citep{Hansen2014} rely on site-level evidence to construct regions.
In addition, several widely used \texttt{R} and Bioconductor packages provide comprehensive workflows for methylation array analysis and incorporate region-finding or downstream DMR functionality, including \textbf{minfi} \citep{Aryee2014}, \textbf{missMethyl} \citep{Phipson2016}, and \textbf{ChAMP} \citep{Tian2017}.
Sequencing-based methylation analysis is supported by packages such as \textbf{methylKit} \citep{Akalin2012}, which similarly construct differentially methylated regions from site-level statistics.
In these approaches, regions are treated as secondary objects derived from CpG-level inference rather than as primary inferential units.

In addition to spatial dependence, methylation measurements often exhibit heterogeneous distributional patterns across samples, including skewness and multimodality, particularly when summarized at the regional level \citep{Irizarry2009, Teschendorff2012}. These features reflect underlying biological heterogeneity and variability across individuals, and they complicate inference procedures that rely primarily on symmetric or mean-based modeling assumptions.

Here we introduce the \texttt{R} \citep{Rbase} package \textbf{mmcmcBayes} \citep{mmcmcBayes2025}, which detects differentially methylated regions using a multistage Markov chain Monte Carlo algorithm. The package is available from the Comprehensive \texttt{R} Archive Network (CRAN) at \url{https://CRAN.R-project.org/package=mmcmcBayes}. The methodology implemented in \textbf{mmcmcBayes} operates on methylation measurements summarized as $M$-values and models region-level methylation distributions separately for the cancer and normal groups, allowing for skewed and bimodal patterns across samples. Evidence for differential methylation is assessed through Bayesian hypothesis testing using Bayes factors, which provide an interpretable measure of evidence while avoiding reliance on permutation-based $p$-values.

A distinguishing feature of \textbf{mmcmcBayes} is its multistage procedure, which begins with broad genomic regions and progressively refines them into more localized segments when supported by statistical evidence. This strategy enables localization of differential methylation signals without requiring prespecified window sizes. The package allows users to specify key tuning parameters, including the number of stages, region-splitting behavior, and Bayes factor thresholds, enabling flexibility across different data types and study designs.

In addition to the main detection procedure, \textbf{mmcmcBayes} provides supporting functions for summarizing, comparing, and visualizing detected regions. The function \texttt{summarize\_dmrs()} reports summary information on identified DMRs by chromosome and stage, \texttt{compare\_dmrs()} facilitates comparisons across analyses or parameter settings, and \texttt{plot\_dmr\_region()} visualizes region-level methylation patterns by plotting mean $M$-values across CpG sites for the cancer and normal groups. The proposed methodology is applicable to both simulated and experimentally generated methylation data and is designed for use with array-based measurements such as those produced by the Illumina HumanMethylation platforms.

The remainder of the paper is organized as follows. Section \ref{sec:models} describes the modeling framework and the multistage Bayesian procedure underlying \textbf{mmcmcBayes}. Section \ref{sec:implementation} presents implementation details of the package, including data structure and primary functions. Section \ref{sec:illustrations} illustrates the use of the \textbf{mmcmcBayes} package through simulation studies and applications to real methylation data. Section \ref{sec:conclusion} concludes the paper with a brief discussion.

\section{Models} \label{sec:models}

\subsection{Alpha-skew Generalized Normal Distribution}\label{sec:asgn_dist}

DNA methylation levels at CpG sites are commonly summarized using $\beta$-values, defined as the proportion of methylated signal relative to the total methylated and unmethylated signal intensity. $\beta$-values lie in the interval $[0,1]$ and provide a natural interpretation as methylation proportions. In this work, methylation measurements are summarized at the region level. For each subject $j$, group $m$, genomic segment $k$, and stage $\ell$, let $\beta_{jml}^\ell$ denote the mean $\beta$-value across all CpG sites within the corresponding segment. Throughout the paper, $m=0$ denotes the cancer group and $m=1$ denotes the control group.

However, direct modeling of $\beta$-values is challenging due to their bounded support. Following standard practice in methylation analysis as \citet{Ryu2013}, $\beta$-values are therefore transformed to $M$-values using logit transformation. Specifically, the $M$-value corresponding to $\beta_{jmk}^\ell$ is defined as 
\begin{equation}
    \label{eq:M value}
        y_{jmk}^\ell = \log\left(\frac{\beta_{jmk}^\ell + c}{1-\beta_{jmk}^\ell + c}\right), 
\end{equation}
where $c>0$ is a small offset introduced to avoid numerical instability where $\beta$-values are close to $0$ or $1$. In practice, we set $c = 10^{-6}$ throughout the package, which is sufficiently small to
ensure numerical stability without materially affecting the transformed values. Although the $M$-value transformation addresses some of the limitations of $\beta$-values, the resulting distributions may still display skewness and bimodality.

To accommodates skewed and asymmetric distributions, the \textbf{mmcmcBayes} package models regional $M$-values using the alpha-skew generalized normal (ASGN) distribution proposed by \cite{Mahmoudi19}. The ASGN distribution extends the normal distribution by incorporating a skewness parameter, allowing asymmetric behavior while retaining a Gaussian kernel, which makes it well suitable for modeling regional methylation summaries. For a given genomic segment $k$ at stage $\ell$ and group $m$, the regional $M$-values $y_{jmk}^\ell$ are assumed to be independently distributed with an ASGN distribution such that 
\[
    y_{jmk}^\ell \simind \text{ASGN} (\alpha_{mk}^\ell, \nu_{mk}^\ell, \delta^{2, \ell}_{mk})
\]
with probability density function 
\begin{equation}\label{eq:asgn}
    f\left(y_{jmk}^\ell \vert \alpha_{mk}^\ell, \nu_{mk}^\ell, \delta^{2\ell}_{mk}\right) = 
    \frac{\sqrt{2}\left\{(1-\alpha_{mk}^\ell y_{jmk}^\ell)^2+1\right\}}{4\left\{\Gamma\left(\frac{3}{2}\right)\alpha_{mk}^\ell + \Gamma\left(\frac{1}{2}\right)\right\}}  \exp\left\{-\frac{\left(y_{jmk}^\ell - \nu_{mk}^\ell\right)^2}{2\delta^{2\ell}_{mk}}\right\},
\end{equation}
where $\alpha_{mk}^\ell \in\mathbb{R}$ controls skewness, $\nu_{mk}^\ell\in\mathbb{R}$ is a location parameter, and $\delta_{mk}^{2,\ell}>0$ is a scale parameter. The same modeling assumption is applied to subsegments at subsequent stages of the analysis. For example, for a subsegments $k^\prime$ at stage $\ell+1$, 
\[
    y_{jmk^{\prime}}^{\ell+1} \simind \text{ASGN}(\alpha_{mk^{\prime}}^{\ell+1}, \nu^{\ell+1}_{mk^{\prime}}, \delta_{mk^{\prime}}^{2,\ell+1}).
\]

For the Bayesian inference, we apply the following prior distributions:
\begin{equation}\label{eq:priors}
    \begin{split}
        \alpha_{mk}^\ell        & \sim N(\mu_a, \sigma^2_a), \\
        \nu_{mk}^\ell           & \sim N(\mu_n, \sigma^2_n), \\
        \delta^{2, \ell}_{mk}   & \sim \text{IG}(A_d, B_d),
    \end{split}
\end{equation}
where $N$ and $\text{IG}$ refer to the Normal and inverse-gamma distributions, respectively. The quantities $\mu_a, \sigma^2_a, \mu_n, \sigma^2_n, A_d$, and $B_d$ are the hyperparameters of the corresponding prior distributions.

Within the multistage Markov chain Monte Carlo framework, prior information is updated sequentially across stages. In particular, priors at stage $\ell+1$ are obtained from the posterior mean from stage $\ell$. For example, the prior distribution for the skewness parameter at stage $\ell+1$ is specified as 
\[
    \alpha_{mk}^{\ell+1}  \sim N(\hat{\alpha}^\ell_{mk}, 1),
\]
where $\hat{\alpha}^\ell_{mk}$ denotes the posterior mean from the previous stage $\ell$. The prior variance is fixed at 1 so that information from earlier stages provides mild guidance without dominating inference for smaller subregions.

\subsection{Multistage MCMC Procedure} \label{sec:mmcmc_procedure}

Given the ASGN model for regional $M$-values described in Section \ref{sec:asgn_dist}, we develop a multistage Markov chain Monte Carlo (MCMC) procedure to adaptively refine genomic regions while performing region-level inference. The approach begins with relatively coarse regions and progressively divide them only when statistical evidence appeared. 

Suppose the procedure proceeds through stages $\ell=1,2,\ldots$. At stage $\ell$, the genomic region is partitioned into $k$ segments. For each segment at stage $\ell$, we model the regional methylation levels for the cancer and normal groups using the ASGN distribution as described in Section \ref{sec:asgn_dist}. 
\begin{figure}[h!]
    \begin{center}\begin{tikzpicture}[scale=1.2]
        \draw[-,gray,xscale=1,yscale=1](0,2) -- coordinate (x axis mid) (8,2);
        \foreach \x in {0,8}\draw[xscale=1,yscale=1] (\x,1.93cm) -- (\x,2.07cm);
        \node at (-1,2)   {$\ell=1$};  
        \node at (4,2.2) {\small Segment$_{1,1}$};
        \draw[-,gray,xscale=1,yscale=1](0,1) -- coordinate (x axis mid) (8,1);
        \foreach \x in {0,4,...,8}\draw[xscale=1,yscale=1] (\x,.93cm) -- (\x,1.07cm);
        \node at (-1,1)   {$\ell=2$};  
        \node at (2,1.2) {\small Segment$_{2,1}$};
        \node at (6,1.2) {\small Segment$_{2,2}$};
        \draw[-,gray,xscale=1,yscale=1](0,0) -- coordinate (x axis mid) (8,0);
        \foreach \x in {0,2,...,8}\draw[xscale=1,yscale=1] (\x,-.07cm) -- (\x,.07cm);
        \node at (-1,0)   {$\ell=3$};  
        \node at (1,.2) {\small Segment$_{3,1}$};
        \node at (3,.2) {\small Segment$_{3,2}$};
        \node at (5,.2) {\small Segment$_{3,3}$};
        \node at (7,.2) {\small Segment$_{3,4}$};
        \node at (-1,-1)   {$\vdots$};  
        \node at (4,-1)   {$\vdots$};  
	\end{tikzpicture}\end{center}    
    \caption{Multistage MCMC procedure}
    \label{fig:chp3mcmc}
\end{figure}
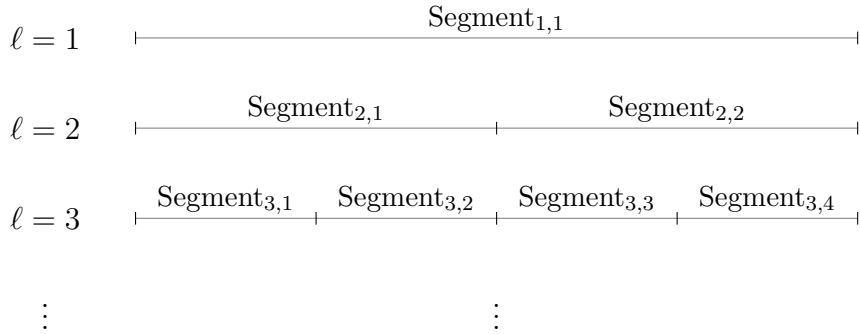
For example, Figure \ref{fig:chp3mcmc} illustrates how the multistage structure is implemented. At the initial stage ($\ell=1$), the entire genomic region is treated as a single segment. For each segment, we quantify evidence for the differential methylation between the two groups using a Bayes factor. When the Bayes factor for a segment exceeds a user-specified threshold, we divide the segment into several subsegments and carry them forward to the next stage ($\ell+1$). When the evidence falls below the threshold, we retain the segment without further splitting. The same procedure is then applied to all segments at the next stage. The procedure continues until a pre-specified maximum number of stages is reached or no segments satisfy the splitting criterion.

\subsection{Model Comparison} \label{sec:bayesfactor}

To formalize region-level comparisons between the cancer and normal groups, we consider two competing models for each genomic segment. Under the null hypothesis $H_0$, regional methylation values from both groups are assumed to arise from a common alpha-skew generalized normal (ASGN) distribution. Under the alternative hypothesis $H_1$, the two groups are modeled using separate ASGN distributions, allowing for differences in their distributional characteristics.

Evidence in favor of differential methylation is quantified using a Bayes factor. Let
$\BF{y}^{\ell}_{0k}$ denote the collection of regional $M$-values for the cancer group and
$\BF{y}^{\ell}_{1k}$ denote the corresponding values for the normal group, both for segment
$k$ at stage $\ell$. The Bayes factor is defined as
\begin{equation*}
BF_{k}^{\ell} =
\frac{
\sum_{j} f\left(y_{j0k}^{\ell}\mid \hat\alpha_{0k}^{\ell}, \hat\nu_{0k}^{\ell}, \hat\delta_{0k}^{2,\ell}\right)
}{
\sum_{j} f\left(y_{j1k}^{\ell}\mid \hat\alpha_{1k}^{\ell}, \hat\nu_{1k}^{\ell}, \hat\delta_{1k}^{2,\ell}\right)
},
\end{equation*}
where $f(\cdot)$ denotes the ASGN density defined in Section~\ref{sec:asgn_dist}, and hats
indicate posterior summary estimates obtained from MCMC sampling. A lower threshold triggers splitting more frequently and tends to produce smaller regions, while a higher threshold triggers splitting less often and tends to produce a larger, coarser regions.

\section{The mmcmcBayes package} \label{sec:implementation}

The \textbf{mmcmcBayes} is available on CRAN. We can use the following code to install and library: 
\begin{Verbatim}[fontsize=\small]
R> install.packages("mmcmcBayes")
R> library("mmcmcBayes")
\end{Verbatim}

\subsection{Data structure} \label{sec:data_structure}

The core function \texttt{mmcmcBayes()} requires two data frames corresponding to the two comparison groups, typically referred as \texttt{cancer\_data} and \texttt{normal\_data}. The two data frames must contain the same CpG sites in the same row order. This requirement is essential because candidate regions are defined and recursively split based on row indices rather than by genomic coordinate matching. As a consequence, the input data must be sorted by genomic position, specifically by chromosome and physical location, prior to analysis. Otherwise, adjacent rows in the input may not correspond to biologically contiguous regions, leading to uninterpretable results. The following code illustrates the structure of the input data by displaying the first five columns of the example dataset distributed with the package:
\begin{Verbatim}[fontsize=\small]
R> data(cancer_demo)
R> head(cancer_demo[,1:5])
    CpG_ID        Chromosome    M_sample_1    M_sample_2    M_sample_3
1  cg00000721          6        1.9781703     1.8551462     2.0203321
2  cg00002930          6       -2.3549588    -1.9455890    -2.0854731
3  cg00003181          6        0.5899144     1.3358503     1.0194316
4  cg00004061          6        1.6567481     0.8350491     0.3173226
5  cg00004562          6        1.1449919     1.1267618     1.0187141
6  cg00004608          6       -0.3366765     0.6255285     0.9428262
\end{Verbatim}

Each row represents information for a single CpG site. The first two columns must be \texttt{CpG\_ID} and \texttt{Chromosome}, which identify the CpG site and its chromosome label. No additional genomic annotation is required. All remaining columns are assumed to contain numeric methylation measurements expressed as $M$-values, as defined in Equation \ref{eq:M value}, for individual samples. For a given region, the \texttt{mmcmcBayes()} function automatically computes the sample wise mean $M$-value across all CpG sites within the region. These region level summaries serve as the observed data in the model and are recomputed at each stage as regions are split during the multistage procedure. Missing values at the CpG level are allowed. When computing region-level means, only available measurements are used.

\subsection{Primary functions} \label{sec:mmcmc_function}

The primary function of the \textbf{mmcmcBayes} package is \texttt{mmcmcBayes()}, which implements a multistage Markov chain Monte Carlo (MCMC) procedure for detecting differentially methylated regions (DMRs) through recursive region splitting, as described in Section \ref{sec:mmcmc_procedure}. The function is loaded as follows:
\begin{Verbatim}[fontsize=\small]
mmcmcBayes(cancer_data, normal_data, 
           stage = 1, max_stages = 3, 
           num_splits = 50, mcmc = NULL,
           priors_cancer = NULL, 
           priors_normal = NULL, 
           bf_thresholds = c(0.5, 0.8, 1.05))
\end{Verbatim}

\begin{itemize}
    \item The argument \texttt{cancer\_data} and \texttt{normal\_data} are data frames containing methylation measurements for the two comparison groups. The required structure of these inputs is described in Section \ref{sec:data_structure}. 
    \item The argument \texttt{stage} specifies the starting stage of the multistage procedure and is typically left at its default value of 1. The maximum number of stages for recursive splitting is controlled by \texttt{max\_stages}. When the procedure reaches \texttt{max\_stages}, or when no subregions satisfy the splitting criterion, no further subdivision is performed. Based on the simulation study presented in Section \ref{sec:simulations}, the default value \texttt{max\_stages = 3} provides a favorable balance between false discovery rate (FDR) and detection precision.
    \item The argument \texttt{num\_splits} controls the number of subregions created when a region is splitting. In practice, the actual number of subregions is bounded by the number of CpG sites within the region. Larger values of \texttt{num\_splits} allow finer localization of DMRs but increase computational cost. The default value \texttt{num\_splits=50} is motivated by simulation results in Section \ref{sec:simulations}, which jointly consider FDR, precision, and computational time.
    \item Users can specify the MCMC sampling setting based on their preference. The argument \texttt{mcmc} is a list containing \texttt{nburn}, \texttt{niter}, and \texttt{thin}. If \texttt{mcmc=NULL}, the default setting \texttt{list(nburn = 5000, niter = 10000, thin = 1)} will be passed directly to the function \texttt{asgn\_func()} to obtain posterior means of the alpha-skew generalized normal (ASGN) model parameters.  
    \item The arguments \texttt{priors\_cancer} and \texttt{priors\_normal} allow users to specify the hyperparameters for the ASGN model, as defined in Equation \ref{eq:priors}, for each group. For example, when $\mu_a=0$, $\mu_n= 2$, and $A_d=1$ for the cancer group, the corresponding prior can be specified as below:
\begin{Verbatim}[fontsize=\small]
R> prior_cancer <- list(alpha=0, mu=2, sigma2=1)    
\end{Verbatim}
    The remaining hyperparameters $\sigma_a^2$, $\sigma_n^2$, and $B_d$ are fixed internally. If the arguments are set to \texttt{NULL}, weakly informative priors are constructed automatically from the observed data at the current stage. 
    \item Stage-specific Bayes factor decision thresholds are specified through \texttt{bf\_threshold}. If the  Bayes factor exceeds the threshold and the current stage is less than \texttt{max\_stages}, the region is split into smaller subregions according to \texttt{num\_splits} and passed to the next stage for further evaluation. If the  Bayes factor exceeds the threshold at the final stage, the region is reported as a detected DMR and no further splitting is performed. If \texttt{NULL}, the default values \texttt{bf\_threshold <- list(0.5, 0.8, 1.05)} will be used. 
\end{itemize}

\subsection{Secondary function} \label{sec:asgn_function}

The secondary function in the \textbf{mmcmcBayes} package is \texttt{asgn\_func()}. The responsibility of this function is the parameter estimation under the alpha-skew generalized normal (ASGN) distribution described in Section \ref{sec:asgn_dist}. The function is loaded by: 
\begin{Verbatim}[fontsize=\small]
asgn_func(data, priors = NULL, 
          mcmc = list(nburn = 5000, niter = 10000, thin = 1),
          seed = NULL)
\end{Verbatim}

\begin{itemize}
    \item The argument \texttt{data} is a univariate numeric vector or a one-column matrix. In the package, this input consists of sample-wise regional mean $M$-values computed for a candidate region. 
    \item The argument \texttt{priors} allows users to specify hyperparameters for the ASGN distribution. This follow the same specification as we introduced in Section \ref{sec:mmcmc_function}. Note that when \texttt{asgn\_func()} is called from \texttt{mmcmcBayes()}, posterior mean from a parent region are reused as the priors for its subregions. 
    \item The \texttt{mcmc} argument follows the same specification as introduced in Section \ref{sec:mmcmc_function}. 
    \item The optional \texttt{seed} argument can be used to set a random seed for reproducibility.   
\end{itemize}

The function returns estimated posterior mean of the ASGN distribution together with corresponding 95\% credible intervals. Although its primary role within the package is to model the regional methylation summaries, the function can also be used independently to fit skewed and potentially bimodal data.

\subsection{Other functions} \label{sec:other_functions}

In addition to the main detection function, the package provides several supporting functions for summarizing, comparing, and visualizing detected differentially methylated regions (DMRs). Table~\ref{tab:supporting_functions} summarizes these supporting functions included in the package and their primary purposes.

\begin{table}[!htbp]
    \centering
    \caption{Supporting functions provided in the \pkg{mmcmcBayes} package.}
    \label{tab:supporting_functions}
    \begin{tabular}{llp{9cm}}
    \toprule
    \toprule
    Functions & Purpose & Description \\
    \midrule
    \texttt{summarize\_dmrs()} &
    Summary &
    The function reports the total number of detected regions, summary statistics of region sizes (\texttt{CpG\_Count}) and decision values, and counts of DMRs by chromosome. If a \texttt{Stage} column is present, counts by stage are also reported. \\
    
    \texttt{compare\_dmrs()} &
    Comparison &
    Identifies overlapping DMRs between two result obtained from different methods or parameter settings. For each overlapping, the function reports an overlap percentage. Returns \texttt{NULL} if no valid overlaps are found. \\
    
    \texttt{plot\_dmr\_region()} &
    Visualization &
    Provides a visualization for interpreting detected DMRs from \texttt{mmcmcBayes()} function by illustrating methylation differences between the cancer and normal groups within each region. \\
    \bottomrule
    \bottomrule
    \end{tabular}
\end{table}

\subsection{Comparison with other packages} \label{sec:comparison}

Table~\ref{tab:dmr_comparison} summarizes key methodological differences between \textbf{mmcmcBayes} and several widely used DMR detection packages.
Existing packages operate primarily at the CpG level and rely on frequentist testing procedures, while \pkg{mmcmcBayes} performs inference directly at the region level within a Bayesian framework. This allows \pkg{mmcmcBayes} to account for skewed or multimodal methylation patterns that are not well captured by CpG-level methods based on mean differences. 
\begin{table}[H]
    \centering
    \caption{ Comparison between \textbf{mmcmcBayes} and other existing packages.}
    \label{tab:dmr_comparison}
    \renewcommand{\arraystretch}{1.00}
    \setlength{\tabcolsep}{4pt}
\begin{tabular}{@{} l p{2.7cm} p{2.6cm} p{2.4cm} p{2.4cm} @{}}
    \toprule
    \textbf{Feature} 
    & \textbf{mmcmcBayes} 
    & \textbf{bumphunter} 
    & \textbf{DMRcate} 
    & \textbf{DMRcaller} \\
    \midrule
    Primary inference level 
    & Region-level 
    & CpG-level 
    & CpG-level 
    & CpG-level \\[10pt]
    
    Statistical framework 
    & Bayesian 
    & Frequentist 
    & Frequentist 
    & Frequentist \\[10pt]
    
    Distributional modeling 
    & Yes 
    & No
    & No
    & No \\[10pt]
    
    Handles skewness 
    & Yes 
    & No
    & No
    & No \\[10pt]
    
    Handles bimodality 
    & Yes 
    & No
    & No
    & No \\[10pt]

    Testing
    & Bayes factor
    & Permutation 
    & FDR 
    & FDR \\

    \bottomrule
    \end{tabular}
\end{table}
Overall, \textbf{mmcmcBayes} provides a complementary region-level approach to existing CpG-level methods by targeting distributional differences rather than mean shifts alone.

\section{Illustrations}\label{sec:illustrations}

We begin by using simulation results to explain the choice of default values in \texttt{mmcmcBaeyes()}. We then demonstrate the use of the package with the demo example distributed with the package, followed by an application to a real Illumina 450K methylation dataset. All illustrative examples in this section were conducted using \texttt{R} version 4.5.1 on a standard macOS platform. Note that the results obtained might for different platforms.

\subsection{Simulation} \label{sec:simulations}
We first conduct a simulation study to justify the default value settings used in the \textbf{mmcmcBayes} package. We use real methylation data from chromosome 6 as a baseline to preserve realistic genomic structure, probe density, and spatial correlation patterns commonly observed in Illumina 450K data. 

To introduce technical and biological variability, we add independent Gaussian noise with mean 0 and standard deviation 0.5 to all samples. In each simulation replicate, ten artificial differentially methylated regions (DMRs) are generated into the dataset. For each DMR, the mean methylation shift is randomly selected to be either $1.0$ or $2.0$ relative to the background, representing moderate to strong biological signals. The length of each DMR is randomly sampled from 10, 20, or 50 consecutive CpG sites, with regions constrained to form non-overlapping genomic blocks. 

To assess the impact of regional partitioning, we evaluate six split configurations (20, 30, 40, 50, 80, and 100 splits). The multistage Bayes factor thresholds are fixed at $(0.5, 0.8, 1.05)$. For each configurations, we recorded sensitivity, the number of false discovery rate, and total computational time. 
\begin{figure}[ht]
    \begin{center}
        \includegraphics[width=\textwidth]{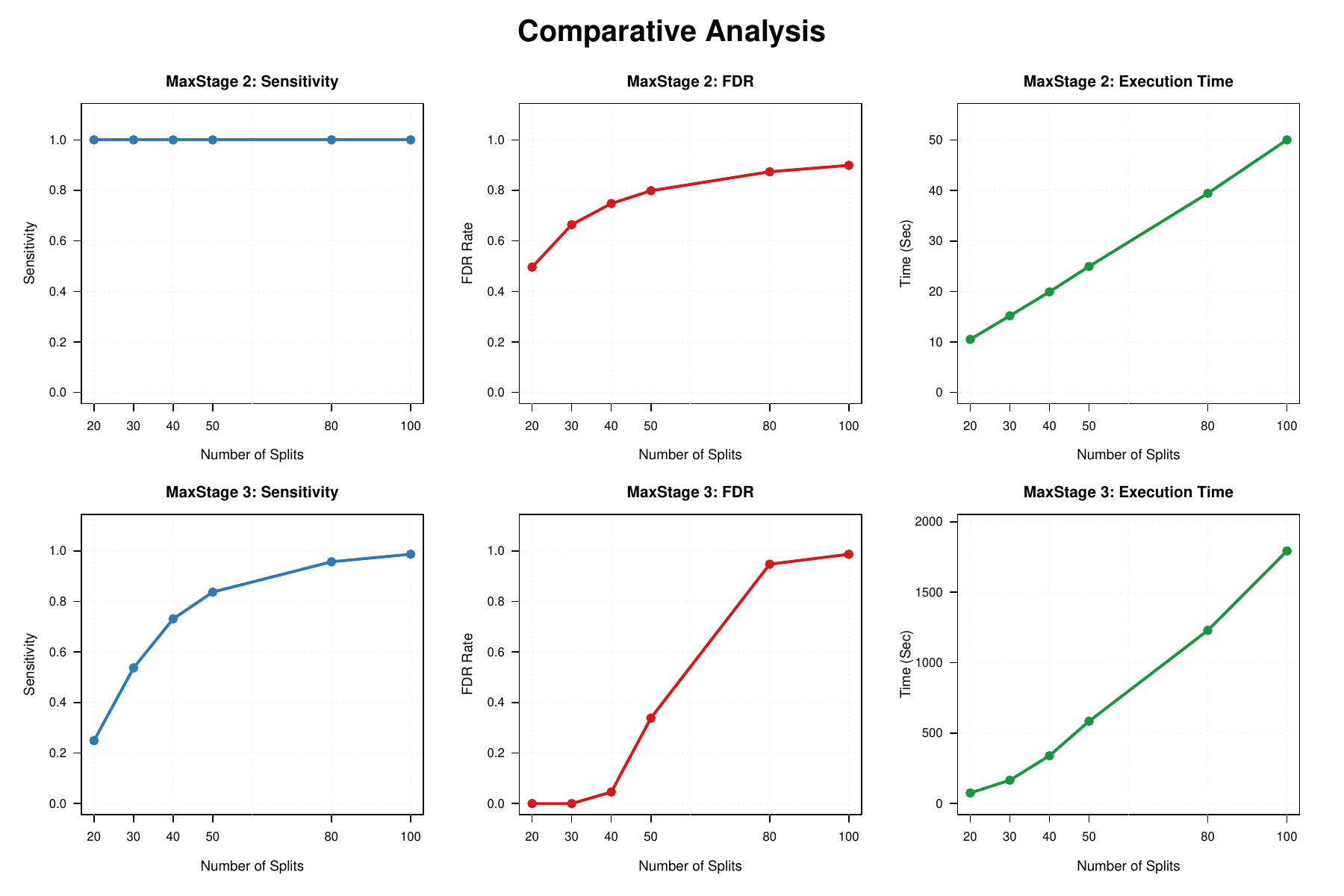}
        \caption{Comparison between MaxStage 2 and MaxStage 3}
        \label{fig:Simulations}
    \end{center}
\end{figure}

Figure~\ref{fig:Simulations} shows how the number of splits affects the results. When \texttt{max\_stage=2}, increasing the number of splits does not improve sensitivity, but it clearly increase the false discovery rate (FDR). When \texttt{max\_stages=3}, sensitivity increases as number of splits increase and reaches a stable level around 50 splits. At the same time, the FDR remains low up to about 50 splits but increases rapidly when more splits are used. Computation time also increases sharply beyond this point. Based on these results, we use \texttt{max\_stage = 3}, \texttt{num\_splits=50}, and the \texttt{bf\_threshold=c(0.5,0.8,1.05)} as the default settings since they give good sensitivity without producing excessive false positives or computation cost. It should be noted that while the package itself runs sequentially, independent analyses across chromosomes can be parallelized by the user using standard \texttt{R} tools.

\subsection{Demo data} \label{sec:demo_data}

In this subsection, we illustrate the basic workflow of \texttt{mmcmcBayes()} using the demo datasets \texttt{cancer\_demo} and \texttt{normal\_demo} included in the package. These datasets contain methylation $M$-values for cancer and normal samples, respectively. Both datasets are derived from the first 5,000 CpG sites on chromosome 6 of an Illumina 450K dataset and are provided as reduced examples for illustration and testing purposes. The following code is used to read the datasets. 
\begin{Verbatim}[fontsize=\small]
R> library("mmcmcBayes")
R> data(cancer_demo, package = "mmcmcBayes")
R> head(cancer_demo[,1:5])

    CpG_ID        Chromosome    M_sample_1    M_sample_2    M_sample_3
1  cg00000721          6        1.9781703     1.8551462     2.0203321
2  cg00002930          6       -2.3549588    -1.9455890    -2.0854731
3  cg00003181          6        0.5899144     1.3358503     1.0194316
4  cg00004061          6        1.6567481     0.8350491     0.3173226
5  cg00004562          6        1.1449919     1.1267618     1.0187141
6  cg00004608          6       -0.3366765     0.6255285     0.9428262

R> data(normal_demo, package = "mmcmcBayes")
R> head(normal_demo[,1:5])

    CpG_ID        Chromosome    M_sample_1    M_sample_2    M_sample_3
1  cg00000721          6        1.9716225     1.93061870    2.2024953
2  cg00002930          6       -2.0804008    -2.42836314   -2.2345472
3  cg00003181          6        0.8372380     0.66398628    0.6713868
4  cg00004061          6        0.7094518     0.38711073    0.4195223
5  cg00004562          6        1.3338421     1.19899281    1.2833079
6  cg00004608          6       -0.1988843    -0.08407308   -0.1059526
\end{Verbatim}

Each dataset is stored as a data frame with CpG sites as rows and sample methylation values as columns. The two datasets share the same CpG ordering and genomic annotations, which allows them to be used directly as paired inputs. The following code is used to run the \texttt{mmcmcBayes()} function. For simplicity, a burn-in of 1,000 iterations is considered, followed by saving 2,000 samples without thinning. 
\begin{Verbatim}[fontsize=\small]
R> set.seed(2021)
R> mcmc <- list(nburn=1000, niter=2000, thin=1)
R> prior_c <- list(alpha=0, mu=2, sigma2=0.1)
R> prior_n <- list(alpha=0, mu=1, sigma2=0.1)
R> rst <- mmcmcBayes(cancer_data = cancer_demo,
+                    normal_data = normal_demo,
+                    stage = 1, max_stages = 2,
+                    num_splits = 5, mcmc = mcmc,
+                    priors_cancer = prior_c,
+                    priors_normal = prior_n,
+                    bf_thresholds = c(0.5, 0.8))
\end{Verbatim}

The arguments \texttt{prior\_cancer} and \texttt{prior\_normal} indicate the hyperparameters defined in Equation \ref{eq:priors}. If \texttt{prior=NULL}, weakly informative priors are automatically constructed from the observed data by the function. 

The output of the returned object \texttt{rst} is given below. 
\begin{Verbatim}[fontsize=\small]
R> print(rst)
    Chromosome  Start_CpG    End_CpG    CpG_Count   Decision_Value  Stage
1       6       cg00000721  cg00613945      1000       9.755850       2
2       6       cg00614280  cg01300096      1000       9.775594       2
3       6       cg01300341  cg01971547      1000       9.732180       2
4       6       cg01971789  cg02683846      1000       9.692494       2
5       6       cg02683893  cg03338154      1000       9.792246       2
\end{Verbatim}

The table listed above summarizes the differentially methylated regions (DMRs) detected by the function. Each row corresponds to one detected region. 
\begin{itemize}
    \item The \texttt{Chromosome} column indicates the chromosome on which the region is located. In this example, all detected regions are on chromosome 6. 
    \item The \texttt{Start\_CpG} column marks the beginning of the detected region. 
    \item The \texttt{End\_CpG} column marks the end of the detected region. 
    \item The \texttt{CpG\_Count} column shows the number of CpG sites contained within the region. 
    \item The \texttt{Decision\_Value} column summarizes the Bayes factor computed for the region. Larger values indicates stronger evidence of differential methylation. 
    \item The \texttt{Stage} column indicates the stage at which the region was identified. For example, a value of 2 means the region satisfied the decision criterion during the second stage of splitting and testing. 
\end{itemize}

We can see that with \texttt{max\_stage=2} and \texttt{num\_splits=5}, the 5,000 CpGs in the demo dataset are examined in coarse blocks of roughly 1,000 CpGs. The regions listed in the output are those that pass the second stage Bayes factor threshold of $0.8$, with \texttt{Decision\_Value} reporting the corresponding Bayes factors. Although these Bayes factors are larger and exceed the second stage threshold, no further splitting is performed because stage 2 is the maximum stage allowed as we set \texttt{max\_stages=2}. 

\subsection{Illumina 450K data} \label{sec:450K_data}

In this subsection, we demonstrate the use of the \textbf{mmcmcBayes} package on a real methylation dataset. The dataset is derived from an Illumina HumanMethylation450 BeadChip (450K) study of lung cancer. It contains methylation measurements from cancer and normal samples restricted to chromosome 6 and in the form of $M$-values. The following code can be used to downloading the dataset from the GitHub repository. 
\begin{Verbatim}[fontsize=\small]
R> c<-"https://raw.githubusercontent.com/zyang1919/mmcmcBayes/main/chr6_cancer.RData"
R> n<-"https://raw.githubusercontent.com/zyang1919/mmcmcBayes/main/chr6_normal.RData"

R> download.file(c, destfile = "chr6_cancer.RData", mode = "wb", quiet = TRUE)
R> download.file(n, destfile = "chr6_normal.RData", mode = "wb", quiet = TRUE)

R> load("chr6_cancer.RData")   
R> load("chr6_normal.RData") 

R> dim(chr6_cancer)

[1] 36438    21

R> dim(chr6_normal)

[1] 36438    21
\end{Verbatim}

The output shows that both datasets contains 36,438 CpG sites and 21 columns. The first two columns store genomic information, \texttt{CpG\_site} and \texttt{Chromosome}, as we showed in Section~\ref{sec:demo_data}, and the remaining 19 columns correspond to sample methylation $M$-values. The cancer and normal datasets have identical dimensions and share the same CpG ordering, allowing them to be used directly as inputs to \texttt{mmcmcBayes()} without additional adjustment. 

The code below applies the \texttt{mmcmcBayes()} function to the loaded cancer and normal datasets. A three stages procedure is used with 50 splits and increasing Bayes factor thresholds across stages, while all other settings are left at their default values. The resulting object is stored as \texttt{rst\_450k}. 
\begin{Verbatim}[fontsize=\small]
R> library("mmcmcBayes")
R> set.seed(2021)
R> rst_450k <- mmcmcBayes(cancer_data = chr6_cancer,
+                        normal_data = chr6_normal,
+                        stage = 1, max_stages = 3,
+                        num_splits = 50,
+                        bf_thresholds = c(0.5, 0.8, 1.05))
\end{Verbatim}

The output obtained by applying the \texttt{summarize\_dmrs()} function from the \textbf{mmcmcBayes} package to the returned object \texttt{rst\_450k} is given below. 
\begin{Verbatim}[fontsize=\small]
R> summarize_dmrs(rst_450k)

Summary of DMR results (mmcmcBayes)
-----------------------------------
Number of DMRs: 1514 
    
Region size (CpG_Count):
   min     q1 median   mean     q3    max 
14.000 14.000 15.000 14.565 15.000 15.000 
    
Decision_Value:
    min     q1    median   mean     q3    max 
    1.050  1.089  1.143    1.184  1.242  1.962 
    
DMRs by Chromosome:
    Chromosome  n_dmrs
1          6     1514
    
DMRs by Stage:
    Stage n_dmrs
1     3   1514
\end{Verbatim}

We can see a total of 1,514 DMRs are detected. Most detected regions are relatively small, with the number of CpG sites per region ranging from 14 to 15 and a median of 15 CpGs. This indicates that the multistage procedure refines candidate regions into narrow segments at the final stage. 

The \texttt{Decision\_Value} column, which corresponds to the Bayes factor for each region, shows that all values exceed the final stage threshold of 1.05, with a median around 1.14. This shows that each reported region meets the strictest decision criterion by the multistage procedure. 

All detected regions are located on chromosome 6, as expected given that the analysis was restricted to this chromosome. In addition, all DMRs are identified at stage 3, which is the final stage of the procedure. This confirms that the procedure first evaluates at large regions and then gradually split them into smaller ones. 

The following code from the \textbf{mmcmcBayes} package plots the detected DMRs on chromosome 6. 
\begin{Verbatim}[fontsize=\small]
R> pdf("Chr6 DMR4 to 9.pdf", width = 11, height = 11)

R> par(mfrow = c(3, 2), mar = c(2.5, 2.5, 2, 1),
+      cex = 0.85, cex.axis = 0.75, cex.lab = 0.85, cex.main = 0.9)

R> plot_dmr_region(rst_450k, chr6_cancer, chr6_normal, dmr_index = 4)
R> plot_dmr_region(rst_450k, chr6_cancer, chr6_normal, dmr_index = 5)
R> plot_dmr_region(rst_450k, chr6_cancer, chr6_normal, dmr_index = 6)
R> plot_dmr_region(rst_450k, chr6_cancer, chr6_normal, dmr_index = 7)
R> plot_dmr_region(rst_450k, chr6_cancer, chr6_normal, dmr_index = 8)
R> plot_dmr_region(rst_450k, chr6_cancer, chr6_normal, dmr_index = 9)
\end{Verbatim}

\begin{figure}[ht!]
    \centering
    { \includegraphics[width=0.8\textwidth]{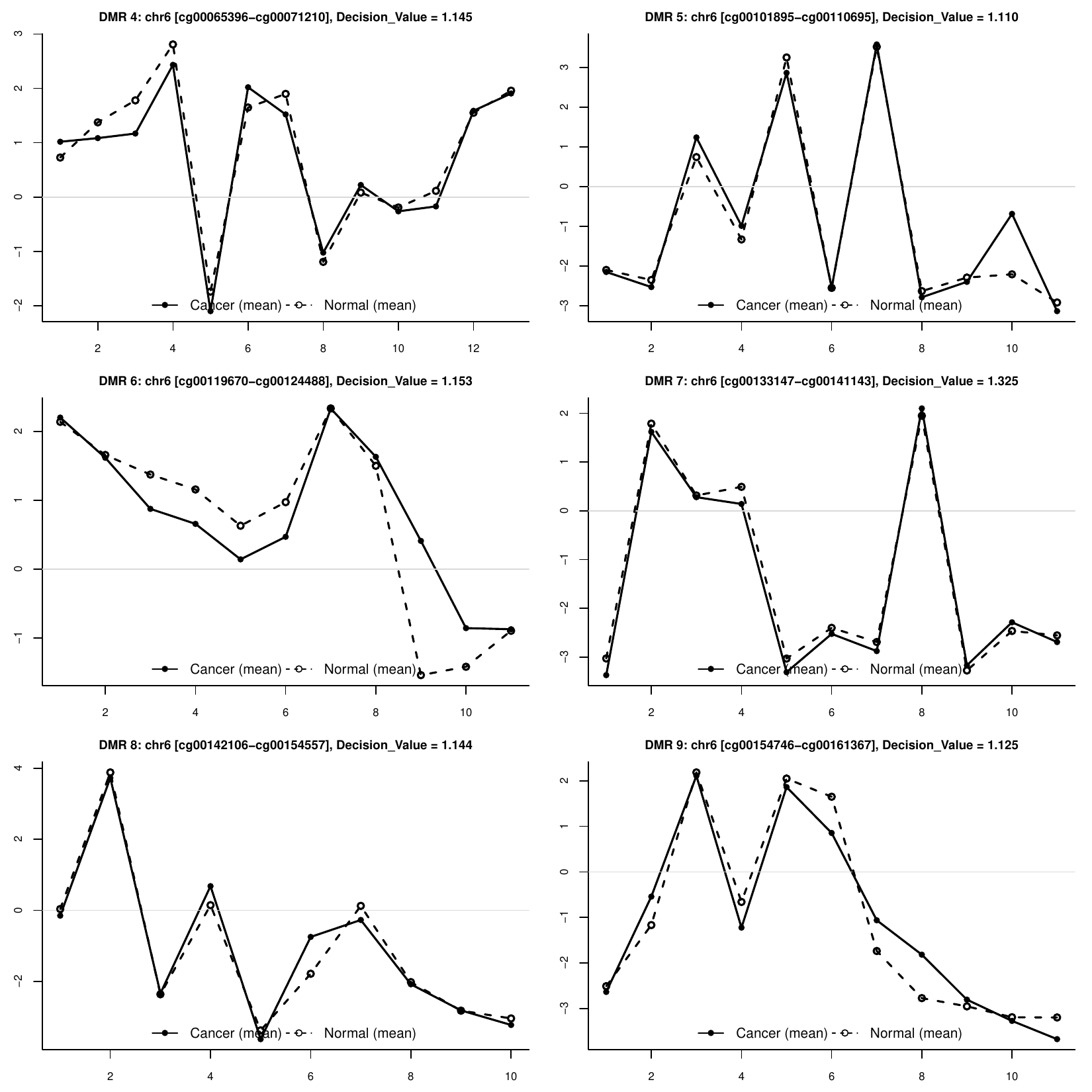}}
    \caption{Example DMRs detected on chromosome 6 in the lung cancer Illumina 450K dataset. Each panel shows the mean $M$-values across CpG sites within a detected region for cancer and normal samples.}
    \label{fig:450k_example_DMR}
\end{figure}

Figure~\ref{fig:450k_example_DMR} plots several representative DMRs detected on chromosome 6 using the function \texttt{plot\_dmr\_region()} from the \textbf{mmcmcBayes} package. Each panel shows the mean $M$-values for cancer and normal samples across the CpG sites within a detected region. The separation between the two groups differs from one to another, reflecting heterogeneity in regional methylation patterns. 

It should be noted that \texttt{plot\_dmr\_regions()} is provided as a visualization tool to evaluate the DMRs detected by \texttt{mmcmcBayes()}. The function requires the DMR result object returned by \texttt{mmcmcBayes()}, the original cancer and normal datasets, and an index specifying which detected region to visualize.

\section{Conclusion} \label{sec:conclusion}

In this paper, we introduced the \texttt{R} package \textbf{mmcmcBayes} for detecting differentially methylated regions (DMRs) using a multistage Markov chain Monte Carlo (MCMC) procedure. The proposed method performs inference directly at the region level by modeling sample-wise regional methylation summaries and comparing cancer and normal groups using Bayes factors. Rather than defining regions as aggregations of CpG-level results, \textbf{mmcmcBayes} treats regions as the primary inferential units and refines their boundaries through a sequence of evidence-based splits.

A key feature of the approach is the use of a multistage MCMC scheme, which begins with coarse genomic segments and progressively focuses on smaller regions only when supported by statistical evidence. This structure provides a natural way to balance localization of differential signals with computational cost. In addition, modeling regional methylation summaries with the alpha-skew generalized normal distribution allows the method to accommodate skewed or asymmetric methylation patterns that are commonly observed in practice but are not explicitly addressed by many existing approaches.

Simulation studies were used to examine the effects of key tuning parameters such as the number of stages and region splits, and to motivate the default settings provided in the package. Applications to both a demonstration dataset and a real Illumina 450K lung cancer dataset illustrate the full analysis workflow and show how the method can be applied to experimentally generated methylation data. Together, these examples demonstrate that \textbf{mmcmcBayes} can identify localized regions of differential methylation while remaining practical to use on moderately large datasets. The \textbf{mmcmcBayes} package provides a reproducible implementation of the proposed methodology, along with tools for summarizing detected regions, comparing results across analyses, and visualizing region-level methylation patterns.

\section*{Acknowledgments}

This article is based on Chapter 3 of the first author's doctoral dissertation \citep{yang2025bfda}, Bayesian Functional Data Analysis and Its Applications.  The authors would like to thank referees for their valuable comments, and all users who have reported bugs and given suggestions.

\bibliographystyle{asa}
\bibliography{refs}

\end{document}